\documentclass[final,5p,times,twocolumn,number,sort&compress]{elsarticle}

\usepackage{amssymb}
\usepackage{lipsum}
\usepackage{amsmath}

\usepackage{xcolor}

\journal{Physics Letters B}

\begin{document}

\begin{frontmatter}


%% use the corref command within \author or \affiliation for footnotes;

\title{Holographic dark energy in a coasting cosmology}

\affiliation[affiliation1]{organization={School of Physics Science and Technology, Xinjiang University},
            city={Urumqi},
            postcode={830017}, 
            country={China}}

\affiliation[affiliation2]{organization={School of Physics and Electronics, Henan University},
            city={kaifeng},
            postcode={475004}, 
            country={China}}

\affiliation[affiliation3]{organization={Xinjiang Astronomical Observatory, Chinese Academy of Sciences},
            addressline={150 Science 1-Street}, 
            city={Urumqi},
            postcode={830011}, 
            country={China}}

\affiliation[affiliation4]{organization={School of Astronomy and Space Science, University of Chinese Academy of Sciences},
            addressline={No.19A Yuquan Road}, 
            city={Beijing},
            postcode={100049}, 
            country={China}}

\affiliation[affiliation5]{organization={School of Physics and Astronomy, China West Normal University},
            city={Nanchong},
            postcode={637002}, 
            country={China}}

\author[affiliation1,affiliation2]{Yunliang Ren}
\author[affiliation2]{Xiaofeng Yang\corref{cor1}}
\author[affiliation3,affiliation4]{Xuwei Zhang}
\author[affiliation1,affiliation2]{Shuangnan Chen}
\author[affiliation2,affiliation5]{Yangjun Shi}
\author[affiliation3,affiliation4]{Cheng Cheng}
\author[affiliation1,affiliation3]{Xiaolong He}
\author[affiliation1]{Hoernisa Iminniyaz\corref{cor2}}

\cortext[cor1]{Corresponding author. Email: xfyang@henu.edu.cn}
\cortext[cor2]{Corresponding author. Email: wrns@xju.edu.cn}

\begin{abstract}

        Coasting cosmology offers an intriguing and straightforward framework for understanding the universe. In this work, 
     we employ the Trans-Planckian Censorship Criterion (TCC) conjecture to test the viability of the coasting cosmology and 
     propose an entropic dark energy (EDE) model within this framework.  
     By applying the holographic principle to constrain the dark energy density and adopting the Bekenstein entropy 
     and Tsallis entropy as the constraining entropies of the system, we find that, 
     in a holographic coasting cosmological framework where dark energy and dark matter evolve independently, 
     the Tsallis entropy satisfies certain general assumptions better than the Bekenstein entropy. 
     Thus, there may be a fundamental relationship between Tsallis entropy and dark energy.
     We utilize observational data from Type Ia Supernovae (SNIa), 
     Baryon Acoustic Oscillations (BAO), and Cosmic Chronometers (CC) to constrain EDE model. 
     The optimal Tsallis parameter obtained aligns well with theoretical expectations. To evaluate the model's fit to the observed data, 
     we calculate the Akaike Information Criterion (AIC), Bayesian Information Criterion (BIC), 
     and Kullback Information Criterion (KIC), and compare these metrics with those derived from $\Lambda$CDM, 
     under which the model shows some improvement. 
     Overall, this model provides a novel and simple on the evolution of the universe.
\end{abstract}

\begin{keyword}

Cosmology \sep Dark energy

\end{keyword}

\end{frontmatter}

\section{Introduction}
\label{introduction}

The origin and evolution of the universe have always been central questions in modern cosmological research. 
The $\Lambda$CDM is based on the FRW space-time, which is also called the standard model of cosmology, and has achieved great success 
in describing the evolution of the universe. The $\Lambda$CDM describes the two episodes of expansion(matter-dominated period and dark energy-dominated period) 
that the universe underwent after the hot Big Bang. Our universe is composed of baryons matter, cold dark matter and dark energy.
After experiencing the radiation-dominated period and the matter-dominated period, the current universe is dominated by dark energy. 
Thus dark energy determines the fate of the universe. Therefore, the dark energy plays a crucial role in the evolution of the universe.

In the $\Lambda$CDM, the density of dark energy is considered to be constant, implying that the expansion of the universe will continue to accelerate. 
In the far-future universe, when dark energy completely dominates the energy composition, the Hubble parameter will approach a constant value, 
leading to perpetual acceleration of the universe's expansion, which contradicts the Trans-Planckian Censorship Criterion (TCC)\cite{TCC1, TCC2, TCC3}. 
In fact, the TCC imposes strict constraints on all cosmological models with eternal accelerated expansion, 
indicating that we need to explore models of dark energy that do not result in eternal expansion in order to avoid inconsistencies in effective field theory. 
Moreover, the latest observations from Dark Energy Spectroscopic Instrument (DESI) suggest that dark energy is likely to be dynamically 
evolving\cite{BAO(DESI1), BAO(DESI2), BAO(DESI3), BAO(DESI4), BAO(DESI5)}, 
which strengthens the case for studying its evolution over time.

Kolb proposed an interesting model named the coasting cosmology\cite{Kolb}. In this model,  the late universe is dominated by K-matter, 
which maintains a uniform expansion of the universe. After this, Chen-Wu type cosmology model implies that a time-varying cosmological constant $\Lambda$ 
as $a^{-2}$, is closer to a constant with certain assumptions in quantum cosmology\cite{chen-wu}. Several challenges in  hot Big Bang models 
might be alleviated in this type of model. In other analogous cosmological models, 
it is evident that the ratio of the density parameters of matter and dark energy consistently behaves as a constant, 
which can effectively avoid the difficulties brought about by inflation\cite{eternal1, eternal2, eternal3, eternal4}. In this type of coasting cosmology model, 
Fulvio Melia also presented a very interesting model, namely the $R_{h}=ct$ universe\cite{FM1, FM2, FM3, FM4, FM5, FM6, FM7, FM8, FM9, FM10}. 
The $R_{h}=ct$ universe can be described as $\Lambda$CDM 
with the additional constraint $w_{tot} = -\frac{1}{3}$, $w_{tot}$ as the total equation of state parameter. Moreover, in this model, 
the various components of the universe do not evolve independently, so there is no need to know the detailed composition of the cosmic fluid. 
Compared to the $\Lambda$CDM model, these models ,which imply that the universe is expanding at a constant rate. 
The merit of $R_{h}=ct$ universe is simple.

On the other side, Geraed't Hooft proposed the famous holographic principle\cite{hooft}, which has significant implications for physics and 
was later applied in the field of dark energy. The original holographic dark energy model is constrained by Bekenstein entropy, 
using the future event horizon as an infrared cutoff, and its dark energy equation of state parameter manifests as a constant determined 
by $c$($c$ is a constant)\cite{holographic1, holographic2, holographic3}. This model describes the acceleration of the expansion of the universe, 
and it has attracted a lot of attention. 
On this basis, physicists have attempted to consider other cutoffs and modified entropies to describe the expansion evolution of the universe, 
aiming to achieve more accurate results\cite{otherHED1, otherHED2,  otherHED3, otherHED4,  otherHED5, otherHED6,  otherHED7,  otherHED8,  otherHED9, 
otherHED10, otherHED11, otherHED12}.

Currently, while some observational data support coasting cosmology, the exploration of its underlying dynamical framework remains an open avenue. 
If we consider the possibility of a simple and uniform expansion of the universe, 
it is essential to establish a consistent theoretical mechanism to explain it.  
In this work, we propose a new dark energy model from entropy in conjunction with the holographic principle. 
This model uses the Hubble horizon as an infrared cutoff and employs both Bekenstein entropy 
and Tsallis entropy as the system's entropy to investigate the evolution of a flat FRW universe without interactions between 
dark energy and dark matter. We constrained the parameters of the dark energy model 
using three types of cosmological observation datasets, 
and obtained the best-fit parameters consistent with the model's expectations. 
Compared with previous studies, our work focuses on the dynamical evolution of holographic dark energy in a coasting cosmology background, 
analyzing its behavior during epochs accessible to current observations, and providing a self-consistent theoretical explanation.
Additionally, we compared the AIC, BIC, and KIC of our model 
with those of the $\Lambda$CDM model to assess the fit of the models to the observational data. 
Finally, we present the conclusions and future directions of this study.

\section{The TCC constraints in coasting cosmology}

The TCC requires that in an expanding universe, 
wavelengths must not exceed the Hubble radius in order to prevent classicalization, which is expressed as:
\begin{equation} \label{TCC}
     \frac{a(t)}{a(t_i)}l_{pl} < H^{-1}(t),
\end{equation}
where $a(t_i)$ is the scale factor at initial time $t_i$, $l_{pl}$ is the Planck length, 
and $H(t)$ is the Hubble parameter (so $H^{-1}(t)$ is the Hubble radius). In the $\Lambda$CDM, 
when the density of dark energy far exceeds that of matter and radiation, the Hubble parameter approaches a constant and no longer varies with time. 
Both $a(t_i)$ and $l_{pl}$ on the left side of the Eq.(\ref{TCC}) are constants, and $a(t)$ increases as the universe expands. 
At a sufficiently late time, the left-hand side will exceed the right-hand side, 
meaning that the $\Lambda$CDM model eventually violates the TCC in the far future, 
indicating that a perpetually accelerating universe is incompatible with the TCC. 
Trans-Planckian modes would ultimately classicalize into the low-energy effective field theory.

In  flat coasting cosmology, the relationship between the critical density $\rho_c$ and the scale factor can be expressed as:
\begin{equation} \label{coasting critical density}
     \rho_{c} = \rho_{c_0} a^{-2},
\end{equation}
and in the far future of the universe, the critical density is replaced by the dark energy density, and the Hubble parameter behaves as:
\begin{equation} \label{far future Hubble parameter}
     H = \frac{\dot{a}}{a} = \frac{\sqrt{\frac{8 \pi \rho_{\Lambda_{0}}}{3}}}{a}.
\end{equation}
In an expanding universe, the scale factor $a$ increases with time, and the Hubble parameter here avoids being treated as a constant. 
From Eq.(\ref{far future Hubble parameter}), we obtain:
\begin{equation} \label{dot scale factor}
     \dot{a} = \sqrt{\frac{8 \pi \rho_{\Lambda_{0}}}{3}},
\end{equation}
\begin{equation} \label{scale factor}
     a = \sqrt{\frac{8 \pi \rho_{\Lambda_{0}}}{3}} t + C,
\end{equation}
 where $C$ is integration constant. We can define the initial time as $t_i$, with the scale factor $a(t_i)$. 
 In this way, the integration constant can be obtained as $C = a(t_i) - \sqrt{\frac{8 \pi \rho_{\Lambda_{0}}}{3}} t_i$. 
Combining with Eq.(\ref{far future Hubble parameter}), we can obtain the Hubble radius: 
\begin{equation} \label{Hubble radius}
     H^{-1} = \frac{\sqrt{\frac{8 \pi \rho_{\Lambda_{0}}}{3}} (t - t_i) + a(t_i)}{\sqrt{\frac{8 \pi \rho_{\Lambda_{0}}}{3}}},
\end{equation}
substituting it into Eq.(\ref{TCC}), we obtain the TCC constraint in the gliding universe, as written:
\begin{equation} \label{coasting TCC}
     \left[1 + \frac{\sqrt{\frac{8 \pi \rho_{\Lambda_{0}}}{3}} (t - t_i)}{a(t_i)}\right] l_{pl} < \frac{ \sqrt{\frac{8 \pi \rho_{\Lambda_{0}}}{3}} (t - t_i) + a(t_i)}{ \sqrt{\frac{8 \pi \rho_{\Lambda_{0}}}{3}}},
\end{equation}
and can be simplified as $l_{pl} < \frac{a(t_i)}{\sqrt{\frac{8 \pi \rho_{\Lambda_{0}}}{3}}}$. We can take the initial time $t_i$ to be the Planck time. 
After substituting the numerical values, it is easy to verify that this inequality holds.

\section{Theoretical model}

The holographic principle indicates that the entropy of a system containing gravity depends on the surface area rather than the volume. 
Based on this assumption, Cohen proposed a relationship between infrared(IR) cutoff and ultraviolet(UR) cutoff\cite{cohen}, 
which can be used to solve the cosmological constant problem, and can be written as :
\begin{equation} \label{IR and UR}
     {L^{3}} \cdot {\Lambda^{3}} \leq {S_{max}} \cong {S^{\frac{3}{4}}_{BH}},
\end{equation}
where $L$ and $\Lambda$ are the IR cutoff and UR cutoff of the system, respectively. The maximum energy density $\Lambda^{4}$ of the system can be 
regarded as vacuum energy, constrained by black hole entropy. 

Cohen proposed a stricter system entropy to relate infrared(IR) cutoff and ultraviolet(UV) cutoff
(${L^{3}} \cdot {\Lambda^{3}} \leq {S_{max}} \cong {S^{\frac{3}{4}}}$, $S$ is black hole entropy, $\Lambda^{4}$ is maximum energy density), 
and the standard holographic dark energy density can be derived as:
\begin{equation} \label{standard holographic dark energy density}
     \rho_{\Lambda} = 3c^{2}M^{2}_{P}L^{-2},
\end{equation}
where $c$ is a numerical constant. In this flat FRW universe, we adopt Hubble horizon($L=H^{-1}$) as the IR cutoff, and Eq.(\ref{standard holographic dark energy density}) 
is expressed as $\rho_{\Lambda} = 3c^{2}M^{2}_{P}H^{2}$. Based on the Friedmann equation, we can obtain the matter energy density,
\begin{equation} \label{standard holographic matter energy density}
     \rho_{m} = 3(1-c^{2})M^{2}_{P}L^{-2}.
\end{equation}  \par
 However, the total energy density $\rho_{tot} \propto a^{-2}$ in coasting cosmology. Thus, the dark energy density 
 and the matter energy density have same 
 exponential relationship with the scale factor $a$ as
 \begin{equation} \label{eternal coasting dark energy density}
     \rho_{\Lambda} \propto a^{-2},
\end{equation}
\begin{equation} \label{eternal coasting matter energy density}
     \rho_{m} \propto a^{-2}.
\end{equation}
It is worth emphasizing that the exponential scaling of the energy densities 
in Eqs.~(\ref{eternal coasting dark energy density})-(\ref{eternal coasting matter energy density}) 
does not arise from an ad hoc identification of the holographic cutoff with the scale factor. 
In standard holographic dark energy models, directly assuming $L \propto a$ is generally considered physically unjustified. 
Instead, in the coasting cosmology considered here, the Hubble radius naturally scales as $H^{-1} \propto a$, 
which leads to the observed scaling behavior when choosing the Hubble horizon as the infrared cutoff ($L = H^{-1}$). 
Therefore, the exponential behavior of $\rho_{\Lambda}$ and $\rho_m$ follows directly from the properties of the coasting background 
and does not introduce any internal inconsistency in the model. 
Therefore, we can conclude that the coasting cosmology dominated by standard holographic dark energy has features similar to 
the quantum coasting cosmology model, with the ratio of matter density to dark energy density being constant. \par

While constructing a cosmological model that reproduces the standard evolution of pressureless matter, 
$\rho_m \propto a^{-3}$, requiring the gravitational system to obey a consistent thermodynamic structure 
naturally motivates going beyond the standard Bekenstein-Hawking entropy. 
It should be emphasized that the Bekenstein-Hawking entropy corresponds to 
an entropic index $\delta = 1$ and obeys the area law. For gravitational systems with long-range interactions, 
this leads to subextensive behavior and consequently violates the Legendre transform structure of thermodynamics. 
As systematically demonstrated in Ref\cite{Tsallis}, when the entropic index takes the value $\delta = 3/2$, 
the entropy scaling $S \propto A^{3/2}$ restores thermodynamic extensivity and respects the Legendre structure. 
Remarkably, this choice is not only theoretically self-consistent but is also supported by recent observational analyses\cite{plb2025}. 
Moreover, an alternative entropic functional that is composable 
but not of trace form has been proposed as a viable possibility alongside $S_\delta$. 
Therefore, in order to recover the classical matter scaling while maintaining thermodynamic consistency, 
we adopt the Tsallis generalized entropy as a representative choice to constrain the system\cite{Tsallis}, whose explicit form is given by:

\begin{equation} \label{Tsallis black hole entropy}
     S_{\delta} = \gamma A^{\delta},
\end{equation}
where $\gamma$ is an unknown constant and $\delta$ is the non-additivity parameter, and $A$ denotes the area of the horizon as $A=4\pi L^{2}$.
Combining Eq.(\ref{standard holographic dark energy density}), the Tsallis holographic dark energy is obtained ,which is 
\begin{equation} \label{Tsallis holographic dark energy density}
     \rho_{\Lambda} = \gamma (4\pi)^{\delta} L^{2\delta - 4} = BH^{4 - 2\delta},
\end{equation}
where $B=\gamma(4\pi)^{\delta}$ denotes an unknown parameter ,and $L$ is Hubble horizon as the IR. We consider that there is no interaction 
between the various components in the universe. Therefore, the dark energy density and the matter energy density follow from conservation law as 
\begin{equation} \label{dark energy conservation law}
     \dot{\rho_{\Lambda}} + 3H\rho_{\Lambda}(1+w_{\Lambda}) = 0,
\end{equation}
\begin{equation} \label{matter conservation law}
     \dot{\rho_{m}} + 3H\rho_{m} = 0,
\end{equation}
where $w_{\Lambda}$ is the equation of state parameter. It is important to clarify that our focus is on the 
late-time coasting universe dominated by dark energy. Similar to the classical coasting cosmology scenario, 
we consider a cosmic history that undergoes standard radiation-dominated ($\rho_r \propto a^{-4}$) and matter-dominated ($\rho_m \propto a^{-3}$) epochs.
Subsequently, as the dark energy component becomes dominant, the universe enters a coasting phase. 
To characterize this specific late-time regime, we adopt the linear expansion ansatz:
\begin{equation}\label{Ez=1+z}
     E(z) = \frac{H(z)}{H_0} = 1 + z,
\end{equation}
which corresponds to a total equation of state $w_{tot} = -1/3$. 
We neglect the existence of the radiation component in the universe dominated by dark energy, without losing much generality.Thus, the total pressure 
is provided by dark energy and matter, which is defined as 
\begin{equation} \label{total pressure}
     P_{tot} = P_{\Lambda} + P_{m},
\end{equation}
where the matter pressure $P_{m} = w_{m} \cdot \rho_{m} = 0$. The matter in the universe is pressureless, and the pressure is provided 
by dark energy.\par
We consider a flat FRW universe made of dark energy and matter, the spatial curvature constant $k=0$. Therefore, the first Friedmann equation is 
written as 
\begin{equation} \label{first Friedmann equation}
     H^{2} = \frac{1}{3M^{2}_{P}}(\rho_{m} + \rho_{\Lambda}), 
\end{equation}
which implies the relationship between the density of the individual components and the evolution of the universe. To facilitate calculation, 
we need define the dimensionless density parameters as 
\begin{equation} \label{dark energy component}
     \Omega_{\Lambda} = \frac{\rho_{\Lambda}}{\rho_{c}} = \frac{B}{3M^{2}_{P}H^{2\delta - 2}},
\end{equation}
\begin{equation} \label{matter component}
     \Omega_{m} = \frac{\rho_{m}}{\rho_{c}} = \frac{\rho_{m}}{3M^{2}_{P}H^{2}},
\end{equation}
where $\rho_{c} = 3M^{2}_{P}H^{2}$ denotes the critical energy density, and $\Omega_{\Lambda} + \Omega_{m} = 1$, 
inserting Eqs.(\ref{Ez=1+z}) and (\ref{total pressure}) ,we obtain 
\begin{equation} \label{dark energy equation of state parameter}
     w_{\Lambda} = -\frac{1}{3}(1 + \frac{1-\Omega_{\Lambda}}{\Omega_{\Lambda}}).
\end{equation}
Now, we take the time derivative of Eq.(\ref{first Friedmann equation}), using Eqs.(\ref{dark energy conservation law}) and (\ref{matter conservation law}), 
and we obtain the second Friedmann equation of the following form as 
\begin{equation} \label{second Friedmann equation}
     \frac{\dot{H}}{H^2} = -\frac{2}{3}(1+\Omega_{\Lambda}w_{\Lambda}).
\end{equation}
There is an exponential relationship between the scale factor $a$ and the time $t$, we can take the derivative of the dark energy density parameter 
$\Omega_{\Lambda}$ with respect to $\ln a$ and obtain
\begin{equation} \label{derivative lna}
     \frac{d\Omega_{\Lambda}}{d\ln a} = (2-2\delta)\Omega_{\Lambda}\frac{\dot{H}}{H^{2}}.
\end{equation}
Inserting Eqs.(\ref{dark energy equation of state parameter}) and (\ref{second Friedmann equation}) into Eq.(\ref{derivative lna}), we have 
\begin{equation} \label{dark energy component a and z}
     \Omega_{\Lambda} = C^{\prime}a^{2\delta - 2} = C^{\prime}(1+z)^{2- 2\delta},
\end{equation}
where $C^{\prime}$ is the integration constant, and the relationship between the factor $a$ and the redshift $z$ is $\frac{1}{a} = 1+z$. In fact, 
we find that $C^{\prime}$ is related to the present dark energy density parameter $\Omega_{\Lambda_{0}}$, thus $C^{\prime}<1$ is required.  We don't consider the case of
 $z<0$(corresponding to the future of the universe), hence $\delta > 2$. Substituting Eq.(\ref{dark energy component a and z}) into 
Eq.(\ref{dark energy equation of state parameter}), we have
\begin{equation} \label{dark energy equation of state parameter z}
     w_{\Lambda} = -\frac{1}{3 \Omega_{\Lambda_{0}}}(1+z)^{2\delta - 2}.
\end{equation}
At this point, we obtain an equation of state parameter of dark energy that evolves with redshift. In addition, the deceleration parameter is defined as 
\begin{equation} \label{deceleration parameter}
     q = -1-\frac{\dot{H}}{H^{2}},
\end{equation}
using Eqs.(\ref{dark energy equation of state parameter}) and (\ref{second Friedmann equation}), we get that $q=0$ corresponds to
 a coasting cosmology.\par

In this way, we obtain a coasting universe model constrained by Tsallis entropy. However, we revisit this model, 
where the dark energy density is given by $\rho_{\Lambda} =  BH^{4 - 2\delta}$. Generally, we assume that the Tsallis parameter $\delta$ and the parameter $B$, 
which includes $\delta$, are constants. In the far future of the universe, the Hubble parameter behaves as follows:
\begin{equation} \label{Tsallis coasting Hubble parameter}
     H = \frac{8 \pi}{3} B H^{3 - 2\delta}.
\end{equation}
We find that if both the Tsallis parameter $\delta$ and the constant $B$ are fixed, 
the Hubble parameter $H$ becomes constant in the very late-time universe, where matter density is negligible. 
This results in a similar challenge as faced by the standard cosmological model. Throughout the derivation of this class of models, 
the Tsallis parameter $\delta$ has been consistently treated as constant during temporal differentiation. 
To address this issue, we propose an alternative approach. For computational convenience, the Friedmann equation can be expressed as follows:
\begin{equation} \label{E=1+z first Friedmann equation}
     H^2 = \frac{8 \pi}{3} (\rho_m + \rho_{\Lambda}).
\end{equation}
By defining the critical density as $\rho_c = \rho_m + \rho_\Lambda$, the matter density parameter can be expressed as :
\begin{equation} \label{E=1+z matter component}
     \Omega_m = \frac{\rho_m}{\rho_c} = \rho_m \frac{8 \pi}{3 H^2} .
\end{equation}
Defining $E_z = \frac{H}{H_0}$ to describe the expansion of the universe, and combining Eq.(\ref{Tsallis holographic dark energy density}) 
and Eq.(\ref{E=1+z first Friedmann equation}) while using the Hubble horizon as the infrared cutoff, we obtain:
\begin{equation} \label{E=1+z Ez}
     E^2_z = \frac{8 \pi}{3H^2_0} \cdot \frac{3H^2}{8\pi} \cdot \Omega_m + \frac{8 \pi B H^{4 - 2\delta}}{3H^2_0} .
\end{equation}
In the coasting cosmology, $E_z$ is given as $E_z = 1 + z$. Substituting this into Eq.(\ref{E=1+z Ez}), 
the matter density parameter can be expressed as:
\begin{equation} \label{matter density parameter}
     \Omega_m = 1- \frac{8\pi B}{3} H^{2-2\delta}_0 (1+z)^{2-2\delta} .
\end{equation}
In a flat universe, where $\Omega_m + \Omega_\Lambda = 1$, and combining this with Eq.(\ref{matter density parameter}), 
the dark energy density parameter can be expressed as:
 \begin{equation} \label{E=1+z dark energy density parameter}
     \Omega_{\Lambda} = \frac{8\pi B}{3} H^{2-2\delta}_0 (1+z)^{2-2\delta} .
\end{equation} 
Based on this, and by substituting into Eq.(\ref{dark energy equation of state parameter}), we obtain the equation of state parameter for dark energy as:
 \begin{equation} \label{E=1+z dark energy equation of state parameter}
     w_{\Lambda} = -\frac{(1+z)^{2\delta - 2}H^{2\delta - 2}_0}{8\pi B} .
\end{equation}
\begin{figure*}[ht!]\suppressfloats
     \centering
     \includegraphics [width=8cm]{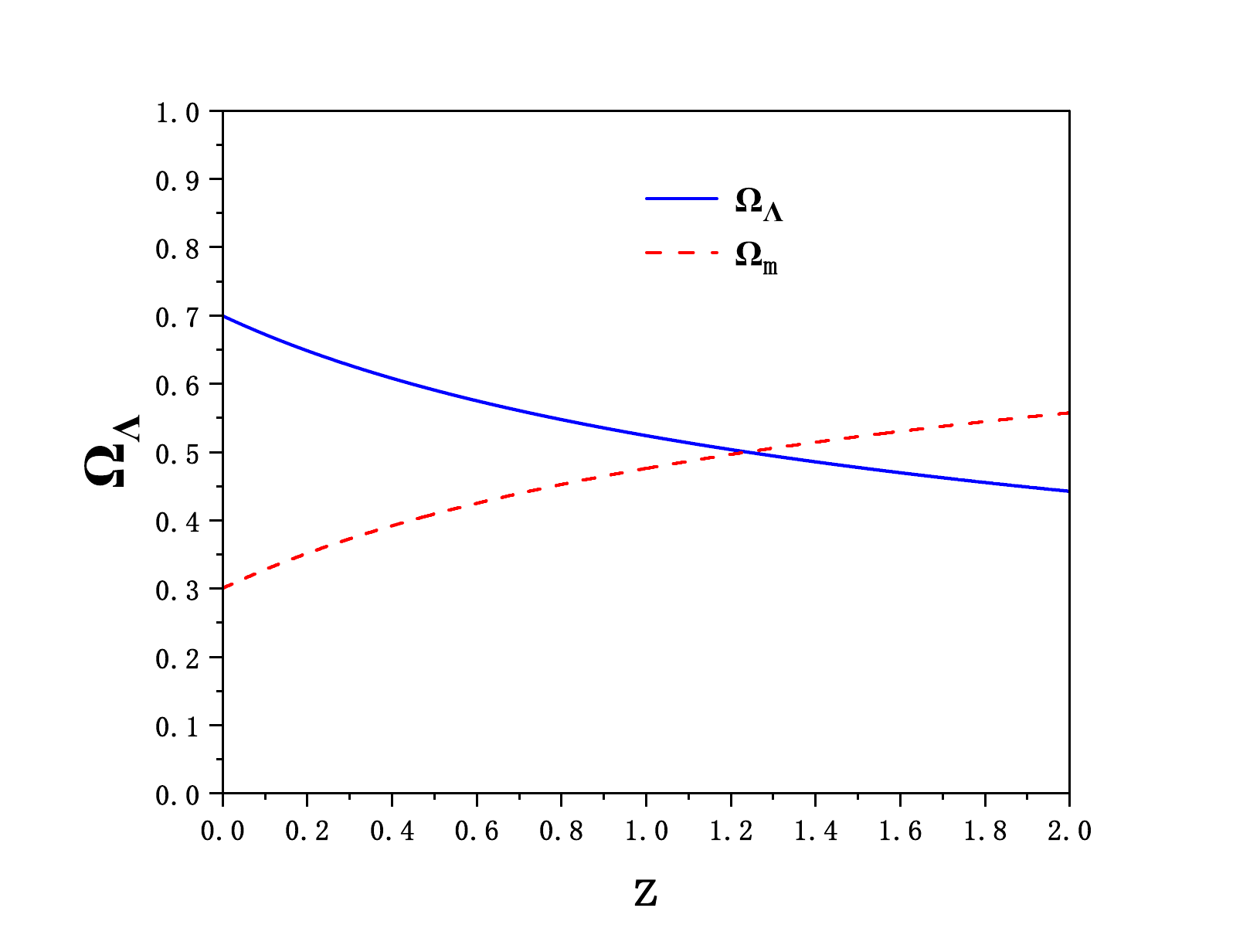}
     \includegraphics [width=8cm]{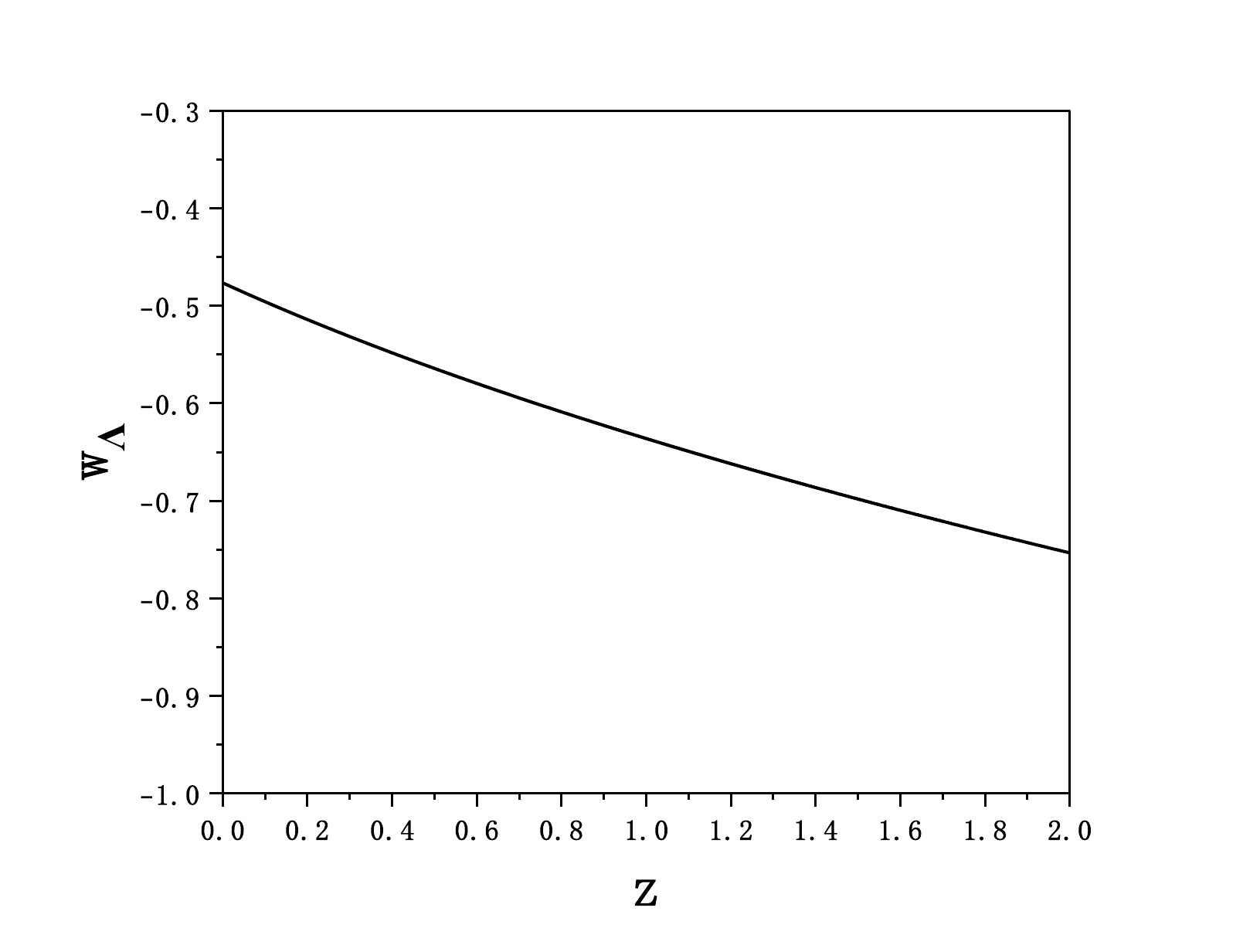}
     \caption{The evolution curves of the dark energy density parameter(left panel) and the equation of state parameter(right panel) as functions of time, 
     under the parameter values $B=0.5$, $\delta = 1.2085$, and $H_0 = 73$(Km/s)/Mpc.}
     \label{Omega and w}
   \end{figure*}

In Figure.\ref{Omega and w}, we present the time evolution of the dark energy density parameter $\Omega_{\Lambda}$ and the equation of state parameter $w_{\Lambda}$ 
in the EDE model under specific parameter values. Compared to the $\Lambda$CDM model, 
this model completes the transition from matter domination to dark energy domination at higher redshifts. 
Moreover, as the redshift decreases, the equation of state parameter $w$ gradually increases, 
indicating that the negative pressure provided by dark energy weakens with the expansion of the universe. 
This allows for a dynamic balance with the gravitational attraction from matter, thereby maintaining a uniform expansion rate.

It is worth noting that, while a dynamically evolving $\delta$ could in principle restore consistency with the 
TCC in the asymptotic future, in Figure \ref{Omega and w} and in the subsequent observational analysis, 
$\delta$ is treated as an effective constant parameter. This choice is motivated by the limitations of current theoretical understanding, 
since a complete quantum gravitational theory deriving the exact time-evolution of $\delta$ is currently unavailable, 
and assuming any specific phenomenological form would therefore be speculative. 
Treating $\delta$ as a constant allows for a robust characterization of the dominant physical behavior during the observationally accessible epoch. 
A systematic exploration of fully time-dependent $\delta$ scenarios is left for future research.

\section{Observational data}

In this work, we analyze and constrain the EDE model using the MCMC method with the emcee Python package\cite{MCMC1, emcee1}. 
The dataset we used includes Type Ia Supernovae, Baryon Acoustic Oscillations, and Cosmic Chronometers, 
allowing us to compare the model with observational data to constrain the model parameters. 
The parameters we need to estimate include $H_0$, $\Omega_m$ and $\delta$.
\begin{figure}[ht!]\suppressfloats
     \centering
     \includegraphics [width=8cm]{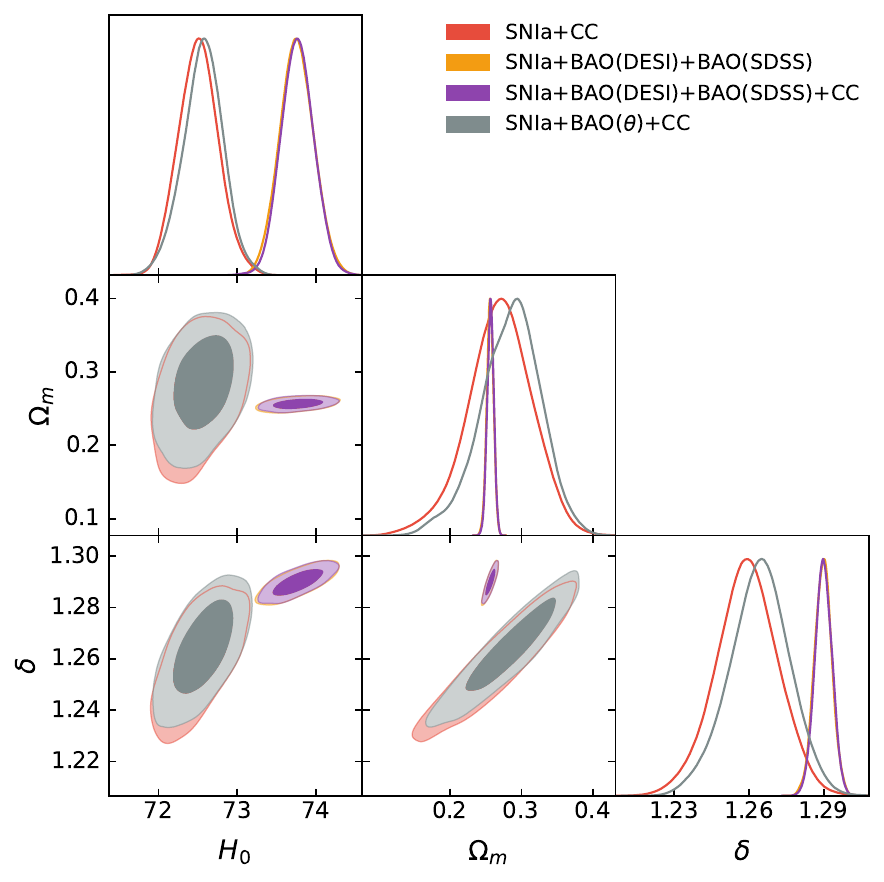}
     \caption{Using different combinations of SNeIa, CC, and three types of BAO data, 
     constrain the 1$\sigma$ and 2$\sigma$ confidence interval distribution plots for the EDE model. 
     The constrained model parameters are $H_0$, $\Omega_{m}$, and $\delta$. 
     Note that the analysis incorporates the full covariance matrices for Pantheon+, standard BAO (SDSS+DESI), 
     and CC data to account for systematic correlations, while the transversal BAO measurements (BAO($\theta$)) are 
     treated with diagonal errors consistent with their statistical independence.}
     \label{coasting-getdist}
   \end{figure}

The SNIa data we used comes from the Pantheon+ sample\cite{pantheon+1}, which contains light curves for 1,550 Type Ia Supernovae 
with a redshift range of $0.001 < z < 2.26$. We define the distance modulus based on the theoretical model as:
\begin{equation} \label{mu}
     \mu_{th}(z) = 5\log_{10} \frac{d_L(z)}{Mpc}+25.
\end{equation}
Here, $d_L(z)$ is the luminosity distance, which can be obtained from $E(z)$ 
in the model. Compared to other probes, SNeIa have more data and are concentrated at lower redshifts.

The BAO data includes BAO($\theta$) (BAO($\theta$) data without assuming a fiducial cosmological model, 
sourced from reference \cite{BAO(NEW)}), BAO (DESI)\cite{BAO(DESI1)}, and BAO (SDSS)\cite{BAO(SDSS)}.
The dark energy parameters can be effectively constrained through the transverse comoving distance $D_M$, 
the Hubble distance $D_H$, and the angular diameter distance $D_V$. Their expressions are as follows:
\begin{equation} \label{DM}
     D_M = \frac{d_L(z)}{1+z}.
\end{equation}
\begin{equation} \label{DH}
     D_H = \frac{c}{H(z)}.
\end{equation}
\begin{equation} \label{DV}
     D_V = (\frac{d_L(z)}{1+z})^{2/3} (\frac{cz}{H(z)})^{1/3}.
\end{equation}

The Cosmic Chronometers(CC) are a model-independent tool for measuring the Hubble parameter, 
which are unaffected by other cosmological assumptions. The CC data used in this paper is sourced from the literature\cite{CC1}, 
which contains 32 data points.

\begin{table*}
\centering     
\begin{tabular}{l c c c} 
 \hline
 Data & $H_0$ & $\Omega_m$ & $\delta$ \\
 \hline
SNIa+CC & $ 72.5097 \pm 0.2549 $ & $ 0.2690 \pm 0.0455 $ & $ 1.2591 \pm 0.0125 $ \\
SNIa+BAO(DESI)+BAO(SDSS) & $ 73.7584 \pm 0.2140 $ & $ 0.2560 \pm 0.0047 $ & $ 1.2896 \pm 0.0034 $ \\
SNIa+BAO(DESI)+BAO(SDSS)+CC  & $ 73.7680 \pm 0.2080 $ & $ 0.2564 \pm 0.0046 $ & $ 1.2898 \pm 0.0034 $ \\
SNIa+BAO($\theta$)+CC  & $ 72.5676 \pm 0.2566 $ & $ 0.2838 \pm 0.0427 $ & $ 1.2641 \pm 0.0121 $ \\
 \hline
\end{tabular}
\caption{The specific values of the best-fitting parameters for Figure \ref{coasting-getdist}.
}
\label{table-best-fitting}
\end{table*}

\begin{table*}
\centering     
\begin{tabular}{l c c c} 
 \hline
 Modle & AIC &  BIC &  KIC \\
 \hline
$\text{EDE}_{\text{SNIa+CC}}$ & 1773.21 & 1789.59 & 1776.21 \\
$\Lambda \text{CDM}_{\text{SNIa+CC}}$ & 1780.15 & 1791.06 & 1782.15 \\
\hline
$\text{EDE}_{\text{SNIa+DESI+SDSS}}$ & 1926.64 & 1942.99 & 1929.64 \\
$\Lambda \text{CDM}_{\text{SNIa+DESI+SDSS}}$ & 1930.94 & 1941.84 & 1932.94 \\
\hline
$\text{EDE}_{\text{SNIa+DESI+SDSS+CC}}$ & 1944.84 & 1961.27 & 1947.84 \\
$\Lambda \text{CDM}_{\text{SNIa+DESI+SDSS+CC}}$ & 1948.93 & 1959.89 & 1950.93 \\
\hline
$\text{EDE}_{\text{SNIa+BAO($\theta$)+CC}}$ & 1788.13 & 1804.53 & 1791.13 \\
$\Lambda \text{CDM}_{\text{SNIa+BAO($\theta$)+CC}}$ & 1794.16 & 1805.10 & 1796.16 \\
\hline
\end{tabular}
\caption{The AIC, BIC, and KIC under different combinations of SNIa+CC and the three BAO data sets, compared with the $\Lambda$CDM model.
}
\label{table-AICBICKIC}
\end{table*}

By jointly constraining the dark energy model using three different cosmological data sets, we obtain the best-fit cosmological parameters for the EDE model.  
Under the joint constraints of various cosmological data sets, the best-fit values of the Tsallis parameter $\delta$ in the model consistently remain below 2, 
in alignment with theoretical expectations. The value of $\delta < 2$ implies that 
the dark energy density decreases with the increasing scale factor of the universe, 
the characteristic that may provide theoretical insights into the fundamental origin of dark energy. 
In comparison with the standard cosmological model, the dark energy in the coasting cosmology exerts a relatively lower negative pressure.

Our observational analysis yields a best-fit value of $\delta \approx 1.28$. 
It is worth noting that, within the framework of generalized thermodynamics, 
the theoretical value $\delta = 3/2$ is typically associated with the restoration of entropy extensivity in a microcanonical ensemble. 
By contrast, the parameter range $1 < \delta < 1.5$ obtained in our analysis may indicate that the effective thermodynamic description 
of the cosmological horizon departs from the ideal microcanonical limit, which could correspond to some form of non-microcanonical (e.g., canonical) thermodynamic behavior. 
Such a possible ensemble-level deviation is consistent with some recent studies\cite{plb2025}, which, based on various cosmological observations, 
also find a preference for non-extensive entropic parameters deviating from the standard area law.

Table\ref{table-AICBICKIC} present the values of AIC, BIC, 
and KIC for the EDE and $\Lambda$CDM models under different combinations of observational data sets. 
These information criteria provide a quantitative basis for assessing the goodness of fit of the EDE model to the observational data. 
Among the data sets used, CC and angular BAO data exhibit relatively weak model dependence. 
Our analysis reveals that, in the vast majority of cases, the EDE model offers a significantly better fit than the $\Lambda$CDM model, 
with the improvement being particularly pronounced in the low-redshift regime.

\section{Conclusion}
We assume the possibility of uniform expansion in the universe and derive the cosmological model 
in which the Trans-Planckian Censorship Conjecture (TCC) is fully satisfied. Specifically, we demonstrate that in coasting cosmology, 
physical wavelengths always remain below the Hubble radius, thus preventing the classicalization of trans-Planckian modes. 
To account for such uniform expansion, we construct a dynamical dark energy model. According to the holographic principle, 
the Hubble horizon is adopted as the infrared (IR) cutoff to constrain the dark energy density.
We constrain the entire system using Bekenstein entropy and found that the ratio between the dark energy density and matter density 
must be a constant, a feature consistent with the quantum coasting universe. 
Tsallis entropy can serve as an alternative to Bekenstein entropy, allowing the Tsallis parameter to be adjusted according to the characteristics of different systems, 
and it is widely applied in the field of cosmology. We choose to use Tsallis entropy to replace Bekenstein entropy to constrain the system, 
with the infrared cutoff still being the Hubble horizon. To satisfy the characteristics of the coasting universe, 
we use $w = -\frac{1}{3}$ to constrain the entire system. This model introduces only one additional parameter compared to $\Lambda$CDM. However, 
we observe that if the Tsallis parameter is treated as a constant, the Hubble parameter asymptotically approaches a constant value in the future, 
thereby violating the TCC. To resolve this, we propose treating the Tsallis parameter as a dynamical quantity and 
construct a new model with two free parameters that satisfies the TCC condition.
We constrain this extended dark energy model using a combination of SNIa, CC, and three different BAO data sets, 
and obtain the best-fit cosmological parameters for various data combinations. 
The resulting best-fit value of the Tsallis parameter is consistent with theoretical expectations. 
Furthermore, we compute the AIC, BIC, and KIC values for the EDE model and compare them with those of the $\Lambda$CDM model 
to evaluate their relative goodness of fit. 
The results indicate that the EDE model consistently provides a significantly better fit to the observational data in most cases.

Finally, exploring the intrinsic physical mechanisms of the coasting cosmology is essential. In this work, we propose a straightforward approach 
to elucidate its evolutionary features, deliberately focusing on simpler scenarios by neglecting complexities 
such as interactions between dark energy and dark matter or the possibility of a non-flat universe.
These aspects merit further exploration and discussion in future research. 
A central issue in dark energy studies is constraining its density. 
Holographic principle may not be the only applicable framework, making the exploration of alternative constraint methods equally important. 
Perhaps there are alternative approaches, and we need to further investigate this.

\section*{Acknowledgements}

This work was supported by the National SKA Program of China (Grants Nos. 2022SKA0110200 and 2022SKA0110203) 
and National Natural Science Foundation of China (No.12463001). 
This work was also supported by Xiaofeng Yang's ZHISHAN Distinguished Professor startup funding of Henan University.

\bibliographystyle{elsarticle-num} 
\bibliography{plb}

\end{document}